\documentclass[eqsecnum,nofootinbib]{revtex4}
\usepackage{slashed,bbold,amsmath,amssymb,mathtools,wrapfig} 
\usepackage{color,epsfig}
\usepackage{hyperref}

\newcommand{\beq} {\begin{equation}}
\newcommand{\eeq} {\end{equation}}
\newcommand{\beqa} {\begin{eqnarray}}
\newcommand{\eeqa} {\end{eqnarray}}

\newcommand{\ie}{{\it i.e.}}

\newcommand{\as}{{\alpha_s}}
\newcommand{\lqcd}{\Lambda_{QCD}}

\newcommand{\ieps}{i\varepsilon}
\newcommand{\vphi}{\varphi}

\newcommand{\order}[1]{${\cal O}\left(#1 \right)$}

\newcommand{\eq}[1]{(\ref{#1})}

\newcommand{\inv}[1]{\frac{1}{#1}}
\newcommand{\halft}{{\textstyle \frac{1}{2}}}

\newcommand{\quart}{{\textstyle \frac{1}{4}}}

\newcommand{\ket}[1]{\left\vert{#1}\right\rangle}
\newcommand{\bra}[1]{\langle{#1}\vert}

\newcommand{\com}[2]{\left[{#1},{#2}\right]}

\newcommand{\bs}[1]{\boldsymbol{#1}}

\newcommand{\xv}{{\bs{x}}}

\newcommand{\kv}{{\bs{k}}}

\newcommand{\gv}{\bs{\gamma}}

\newcommand{\gz}{\gamma^0}

\newcommand{\nv}{\bs{\nabla}}

\newcommand{\rar}{\rightarrow}
\newcommand{\lar}{\leftarrow}

\newcommand{\rnab}{{\overset{\rar}{\nv}}}
\newcommand{\lnab}{{\overset{\lar}{\nv}}}

\newcommand{\alv}{{\bs{\alpha}}}

\begin{document}

\vspace{-.2cm}
\title{The Born approximation for bound states\footnote{Based on a talk at the XVII International Conference on Hadron Spectroscopy and Structure -- Hadron2017 on 25-29 September, 2017 at the University of Salamanca, Salamanca, Spain. The title of the talk was {\em Hadrons in Born approximation.}}}

\author{Paul Hoyer \vspace{.2cm}} 

\affiliation{Department of Physics, POB 64, FIN-00014 University of Helsinki, Finland}

\begin{abstract}
Bound states are stationary in time and interact continuously. Even a first approximation of atomic wave functions in QED requires contributions of all orders in $\alpha$. Bound state perturbation theory depends on the choice of this first approximation, just as the Taylor expansion of an ordinary function depends on the expansion point. Considering the expansion to be not in $\alpha$ but in $\hbar$, \ie, in the number of loops, defines the perturbative expansion uniquely also for bound states. I show how the Schr\"odinger equation for Positronium with the classical potential $V(r)=-\alpha/r$ corresponds to the Born, \order{\hbar^0} bound state approximation in QED.

Standard perturbation theory is based on an expansion around \order{\alpha^0} free states that have no overlap with bound states. Perturbing around bound states requires using interacting $in$ and $out$ states. For Born states the binding potential arises from a classical gauge field. In the absence of loops the QCD scale $\lqcd$ can originate from a boundary condition imposed on the solution of the classical gluon field equations. A perturbative expansion may be relevant even for hadrons, if their non-perturbative features such as confinement and chiral symmetry breaking are present already in the Born term.
\end{abstract}

\maketitle

\section{The $\hbar$ expansion for bound states}

Scattering amplitudes are typically well described by perturbation theory at energy scales where bound state effects may be neglected (or factorized). The lowest order (Born term) contribution is given by \order{\alpha^n} tree diagrams, where $n$ depends on the number of external legs ($n=1$ for $2 \to 2$ processes). The perturbative expansion is equivalent to a loop expansion, with each loop contributing a factor $\alpha\hbar$. Since the powers of $\alpha$ and $\hbar$ are linked one usually sets $\hbar=1$. 

The powers of $\hbar$ may be viewed as arising from fluctuations in the quantum field. The functional integral measure of a generic field $\vphi$ is
\beq\label{green}
\int [d\vphi]\,\exp\big(i\,S[\vphi]/\hbar\big)
\eeq
where $S[\vphi]$ is the action.
The leading contribution in the $\hbar \to 0$ limit is given by the classical field $\vphi_{cl}$, for which the action is stationary. Fluctuations $\vphi-\vphi_{cl}$ are suppressed by powers of $\hbar$, as are the loops of Feynman diagrams. I shall argue that the $\hbar$ expansion is useful for bound states, and distinct from the expansion in $\alpha$. 

Bound state contributions to a scattering process involve all powers of $\alpha$, even in a first approximation. For example, the formation of Positronium in the direct channel of $e^+e^- \to e^+e^-$ implies a pole in the scattering amplitude,
\begin{align} \label{eescat}
A(e^+e^- \to e^+e^-) \sim \frac{\Phi_i^* \Phi_f}{s-M^2+\ieps}
\end{align}
The Positronium mass is $M = 2m_e+E_b$, where the binding energy is $E_b = -\quart m_e\alpha^2$ for the ground state (at lowest order in $\alpha$). The denominator of \eq{eescat} expands to an infinite series in powers of $\alpha$. Furthermore, the wave functions $\Phi_{i,f}$ which describe the coupling of the Positronium to the initial and final $e^+e^-$ states are exponential functions of $\alpha$.

Bound state perturbation theory depends on the choice of the \order{\alpha^\infty} first approximation. Powers of $\alpha$ may be shifted between the first approximation and the perturbative expansion around it, giving an infinite set of equivalent, formally exact expansions for bound states. This is in stark contrast to scattering amplitudes, where the first approximation (Born term) is unique.

\begin{figure}[h] 
  \begin{minipage}{.5\textwidth}
    \begin{align}\label{bspos}
\Phi(P,k) = \int \frac{d^4q}{(2\pi)^4}\, \Phi(P,k+q)\, S(P,k,q)\, K(P,k,q)
\end{align}
  \end{minipage}%
  \begin{minipage}{.5\textwidth}
    \hfill
    \includegraphics[width=0.9\textwidth]{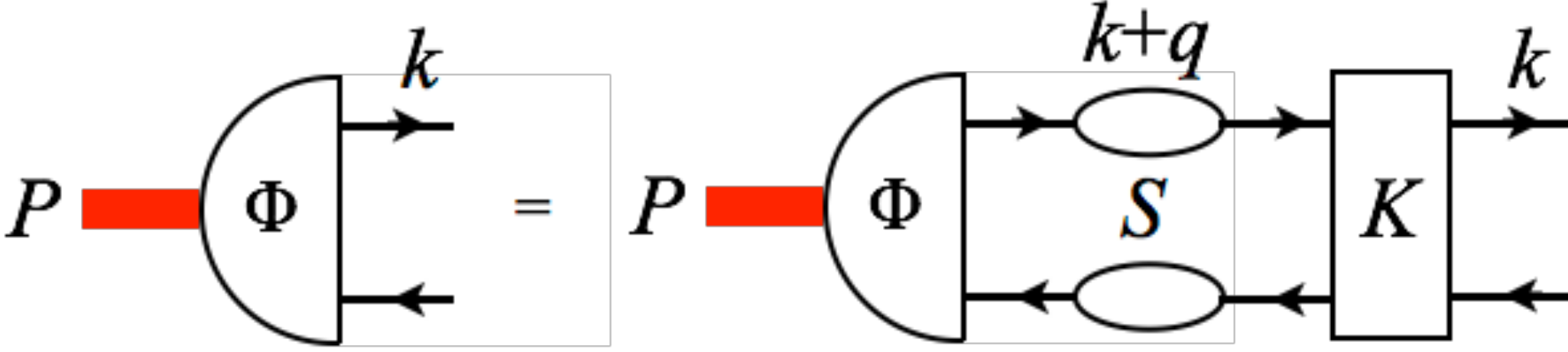}
  \end{minipage}
\end{figure}
The non-uniqueness of bound state perturbation theory may be illustrated by the Bethe-Salpeter equation \eq{bspos} for the Positronium wave function $\Phi$ \cite{Salpeter:1951sz}. The B-S equation is formally exact when the $e^+e^-$ propagator $S$ and the (2-particle irreducible) interaction kernel $K$ are expanded to all orders in $\alpha$.
Caswell and Lepage \cite{Caswell:1978mt} pointed out that the expansion of $K$ depends on the choice of $S$. This allowed for considerable simplifications. The B-S equation \eq{bspos} has no analytic solution even when $S$ is given by free (relativistic) propagators and $K$ has only single photon exchange. Choosing non-relativistic propagators for $S$ the B-S equation reduces to the familiar Schr\"odinger equation at lowest order in $K$, with calculable power corrections in $\alpha$.

Present calculations of atomic spectra make use of the effective theory NRQED, which is based on an expansion of the QED action in powers of $\nv/m$ \cite{Caswell:1985ui}. Applications of NRQED to atoms perturb around a lowest order term defined by the Schr\"odinger equation, which is a practical and successful choice. The purpose of the present note is to point out that this choice is also consistent with a general principle. The Schr\"odinger state has no loop contributions, \ie, it corresponds to an \order{\hbar^0} Born contribution. Since the $\hbar$ expansion is defined also for relativistic dynamics one may consider whether the Born approximation could be a useful starting point even for QCD hadrons.

For atomic kinematics the momenta in $e^+e^- \to e^+e^-$ scale with $\alpha$ as $|{\bf p}| \sim m_e\alpha$ (Bohr momentum) and $p^0 \sim {\bf p}^2/2m_e \sim \halft m_e\alpha^2$. This gives rise to inverse powers of $\alpha$ from the propagators, causing all ``ladder'' diagrams to scale with the same power of $\alpha$. Their sum is a geometric series of the single photon exchange contribution, and has bound state poles in agreement with the Schr\"odinger equation.

The sum of ladder diagrams appears to be a fully perturbative derivation of the Schr\"odinger states. However, it suffers from a flaw of principle which is crucial in applications to QCD. Feynman diagrams are generated by the expression for the $S$-matrix in the Interaction Picture,
\beq\label{smatrix}
S_{fi}={}_{out}\bra{f,t\to \infty}\left\{ {\rm T}\exp\Big[-i\int_{-\infty}^\infty dt\,H_I(t)\Big]\right\}\ket{i,t\to -\infty}_{in}
\eeq
where $\ket{i}_{in}$ and ${}_{out}\bra{f}$ are eigenstates of the free Hamiltonian $H_0$. The interaction Hamiltonian $H_I$ dresses the bare asymptotic states, ensuring that they scatter as physical states. The free $e^+$ and $e^-$ in the $in$ and $out$ states are assumed to be infinitely separated and thus have no overlap with a Positronium state. Hence the expression \eq{smatrix} is, in principle, not applicable to bound states.

Even for Positronium in QED we thus need to choose a first, \order{\alpha^\infty} approximation for the bound state, in place of the \order{\alpha^0} $in$ and $out$ states of \eq{smatrix}. Any asymptotic state that overlaps Positronium is allowed, since states at $t=\pm\infty$ will relax to the physical states during time evolution. The interactions included in the first approximation (Coulomb photon exchange for the Schr\"odinger atom) should be subtracted from those generated by $H_I$ to avoid double counting.

The choice of potential in the Schr\"odinger equation is motivated by $V(r) = -\alpha/r$ being the {\em classical} potential. Recalling the relation \eq{green} between the classical field and the $\hbar \to 0$ limit suggests that the Schr\"odinger atom is the Born term of physical Positronium, similarly as tree diagrams are Born terms of scattering amplitudes. In the next section I demonstrate how the Schr\"odinger equation for Positronium arises from binding by the classical electromagnetic field. The $\hbar$ expansion is a guide in the choice of bound state around which the perturbative series is developed. This may allow a perturbative approach to hadrons in QCD, with the non-perturbative features present already in the Born term.

\section{Positronium in Born approximation \label{positroniumsec}}

Let us consider how the Schr\"odinger equation for Positronium at rest is obtained in the Born approximation of QED. The general bound state condition for a state $\ket{M}$ is
\begin{align} \label{qedbse}
H_{QED}\ket{M}= M\ket{M}
\end{align}
In the non-relativistic limit $\ket{M}$ has only an $\ket{e^+e^-}$ Fock state described by the Schr\"odinger wave function $\phi(\xv)$, which is independent of the $e^\pm$ helicities $\lambda_{1,2}$\,,
\begin{align} \label{stats}
\ket{M} = \int\frac{d\kv}{(2\pi)^3}\,\phi(\kv)\,b^\dag_{\kv,\lambda_1}d^\dag_{-\kv,\lambda_2}\ket{0} = \int d\xv_1\,d\xv_2\,\overline\psi_\alpha(0,\xv_1)\,\Phi_{\alpha\beta}(\xv_1-\xv_2)\,\psi_\beta(0,\xv_2)\ket{0}
\end{align}
Comparing the coefficient of $b^\dag d^\dag$ in the two expressions, given that
\begin{align}
\psi_\alpha(t=0,\xv) = \int\frac{d\kv}{(2\pi)^3 2E_k}\sum_\lambda\big[u_\alpha(\kv,\lambda)e^{i\,\kv\cdot\xv}b_{\kv,\lambda}+v_\alpha(\kv,\lambda)e^{-i\,\kv\cdot\xv}d_{\kv,\lambda}^\dag\big]
\end{align}
allows to express the $4\times 4$ wave function $\Phi(\xv_1-\xv_2)$ in terms of the Schr\"odinger wave function,
\begin{align} \label{dirstr}
\Phi_{\alpha\beta}(\xv) = {_\alpha\big[}\gz u(-i\,\nv,\lambda_1)\big]\big[\bar v(i\,\nv,\lambda_2)\gz\big]_\beta\,\phi(\xv) \hspace{1cm} {\rm where} \hspace{1cm} \phi(\xv) = \int\frac{d\kv}{(2\pi)^3}\phi(\kv)\,e^{i\kv\cdot\xv}
\end{align} 
The Dirac equations satisfied by the $u$ and $v$ spinors ensure that with $E = \sqrt{-\nv^2+m^2}$ and $\alv=\gz\gv$,
\begin{align} \label{projid}
\Big[\inv{E}(i\alv\cdot\rnab+m\gz)-1\Big]\Phi(\xv) = \Phi(\xv)\Big[(i\alv\cdot\lnab-m\gz)\inv{E}-1\Big] = 0
\end{align}  

The Born approximation implies that we use the classical photon field in the Hamiltonian of the bound state condition \eq{qedbse}. We may neglect the 3-vector fields ${A}^j$ due to the non-relativistic dynamics of Positronium (in the rest frame): The ${A}^j$ couple to the 3-momenta $p^j \sim \alpha\, m$ of the fermions, which are negligible compared to $p^0 \simeq m$.

The operator equation of motion for the $\hat A^0$ field does not involve a time derivative,
\beq\label{gausslaw}
\frac{\delta{S}_{QED}}{\delta \hat A^0(t,\xv)}=0 \hspace{1cm} \Rightarrow \hspace{1cm}
-\nv^2 \hat A^0(t,\xv)=e\psi^\dag(t,\xv)\psi(t,\xv)
\eeq
Hence it determines ${\hat A}^0(t,\xv)$ in terms of the electron field at each instant of time, for all positions $\xv$. In gauge theories there are no retardation effects for ${\hat A}^0(t,\xv)$ even when the dynamics is relativistic.

The photon field energy contributes to $H_{QED}$ as
\begin{align} \label{fielden}
\halft\int d\xv\, F_{i0}F^{i0} = \halft\int d\xv\, \hat A^0\,\nv^2 \hat A^0 = -\halft\int d\xv\,\overline{\psi}\,e\gz \hat A^0\psi
\end{align}
which cancels half of the interaction term in the fermion part of the Hamiltonian. Thus
\begin{align} \label{hamrest}
H_{QED}(t) = \int d\xv\,\psi^\dag(t,\xv)\big[-i\nv\cdot\alv+m\gz+\halft e\hat{A}^0(\xv)\big]\psi(t,\xv)
\end{align}

The classical photon field for a state where the electron is at $\xv_1$ and the positron at $\xv_2$ is the expectation value of $\hat{A}^0(t,\xv)$ in that state, $\ket{\xv_1,\xv_2} \equiv \psi^\dag(t,\xv_1)\psi(t,\xv_2)\ket{0}$,
\begin{align}
A^0(t,\xv;\xv_1,\xv_2) \equiv \frac{\bra{\xv_1,\xv_2}\,\hat{A}^0(t,\xv)\,\ket{\xv_1,\xv_2}}{\bra{\xv_1,\xv_2}\xv_1,\xv_2\rangle}
\end{align}
Note that the classical field depends on the positions $\xv_1,\,\xv_2$ of the $e^-$ and $e^+$. In discussions of the Schr\"odinger equation one commonly reduces the two-body problem to that of a single particle in a fixed potential. This obscures the notion of the classical field that is used here.

The operator relation \eq{gausslaw} implies the usual Gauss' law for the classical photon field (at $t=0$),
\begin{align} \label{gaussqed}
-\nv^2_x A^0(\xv;\xv_1,\xv_2) = e\big[\delta(\xv-\xv_1)-\delta(\xv-\xv_2)\big]
\end{align}
with the standard solution
\begin{align} \label{coulfield}
eA^0(\xv;\xv_1,\xv_2) = \frac{\alpha}{|\xv-\xv_1|} - \frac{\alpha}{|\xv-\xv_2|}
\end{align}

Having determined the classical field for each configuration $\ket{\xv_1,\xv_2}$ we may impose the bound state condition \eq{qedbse}. We use $H_{QED}\ket{0}=0$ (no pair production in the non-relativistic limit), and
\begin{align}
\com{H}{\overline\psi(\xv_1)} &= \overline\psi(\xv_1)\big[-i\alv\cdot\lnab_1+m\gz+\halft eA^0(\xv_1)\big] \label{hampsib}\\
\com{H}{\psi(\xv_2)} &= -\big[-i\alv\cdot\rnab_2+m\gz+\halft eA^0(\xv_2)\big]\psi(\xv_2)  \label{hampsi}
\end{align}
where the direction of differentiation is indicated by an arrow and we used the short-hand notation $A^0(\xv_1) \equiv A^0(\xv_1;\xv_1,\xv_2)$ for the classical field at $\xv_1$ in the state $\ket{\xv_1,\xv_2}$. According to \eq{coulfield} both $A^0(\xv_1)$ and $A^0(\xv_2)$ contain a ``self-energy'' term $\sim\alpha/0$. They shift the Positronium mass by an (infinite) amount that is independent of $\xv_1,\,\xv_2$ and may be subtracted. Thus we have
\begin{align} \label{potdef}
eA^0(\xv_1;\xv_1,\xv_2)=-eA^0(\xv_2;\xv_1,\xv_2)=-\frac{\alpha}{|\xv_1-\xv_2|} \equiv V(\xv_1-\xv_2)
\end{align}

The bound state condition \eq{qedbse} imposes (after partial integrations) the bound state equation for the wave function $\Phi(\xv_1-\xv_2)$ of the $\ket{\xv_1,\xv_2}$ component of the Positronium state,
\begin{align} \label{nrbse1}
\big[i\alv\cdot\rnab+m\gz+\halft V(\xv)\big]\Phi(\xv)+ \Phi(\xv)\big[i\alv\cdot\lnab-m\gz+\halft V(\xv)\big] = M\Phi(\xv)
\end{align}
with the classical potential $V(\xv)$ defined in \eq{potdef}. If we bring the term $\propto M$ over to the lhs., share it equally between the two terms and divide the equation by $\halft(M-V)$ the BSE becomes
\begin{align} \label{nrbse2}
\Big[\frac{2}{M-V}(i\alv\cdot\rnab+m\gz)-1\Big]\Phi(\xv)+\Phi(\xv)\Big[(i\alv\cdot\lnab-m\gz)\frac{2}{M-V}-1\Big]=0
\end{align}
The potential $V$ as well as the binding energy $E_b \equiv M-2m$ are of \order{\alpha^2} so we may expand,
\begin{align} \label{mvexp}
\frac{2}{M-V} \simeq \inv{E}+\inv{2m^2}\Big[-\frac{\nv^2}{m}+V-E_b\Big]
\end{align}
where we used $E=\sqrt{-\nv^2+m^2} \simeq m-\nv^2/2m$. The constraints \eq{projid} ensure that the term $1/E$ does not contribute in the BSE \eq{nrbse2}. The second term in \eq{mvexp} can be brought past the derivatives in the expression \eq{dirstr} for $\Phi$, since $\nv V(\xv)$ is of \order{\alpha^3} and may be ignored. Both terms in the BSE \eq{nrbse2} then vanish separately since $\phi(\xv)$ satisfies the Schr\"odinger equation (with reduced mass $\halft m$),
\begin{align}
\Big[-\frac{\nv^2}{m}+V\Big]\phi(\xv) = E_b \phi(\xv)
\end{align}

\section{Comments \label{comsec}}

Bound state perturbation theory requires to select an initial, approximative state around which the perturbative expansion is developed. For this state to have a non-vanishing overlap with the physical bound state it must be non-polynomial in $\alpha$. All such states are formally equivalent since the difference in their $\alpha$-dependence may be shifted into the perturbative expansion. The $S$-matrix with bound states in the initial or final states should be determined by an expression analogous to \eq{smatrix}, where an approximative bound state is used in the $in$ and $out$ states at $t=\pm\infty$.

In practice already the lowest order of the perturbative expansion should provide a good approximation of the physical state. The $\hbar$ or loop expansion is successful for scattering amplitudes and at \order{\hbar^0} selects bound states formed by the classical gauge field of their constituents. I illustrated how the Born state of Positronium is defined by the Schr\"odinger equation with the potential arising from the classical photon field $A^0(\xv;\xv_1,\xv_2)$ given in \eq{coulfield}. Since $A^0$ is a solution of the equations of motion it is different for each component $\ket{\xv_1,\xv_2}$ of the bound state.

The sum of ladder diagrams seems like a purely perturbative derivation of the Schr\"odinger atom. However, Feynman diagrams describe perturbative corrections to the non-interacting $in$ and $out$ states of \eq{smatrix} which have no overlap with bound states. In QCD the standard perturbative expansion around free quark and gluon states implies initial configurations with infinitely separated color charges. Establishing confinement from Feynman diagrams is therefore challenging, if not impossible. This does not exclude the possibility of a perturbative expansion around initial bound states that already have confinement. The loop ($\hbar$) expansion provides a principle for the choice of lowest order term. 

In the absence of loops $\as$ is frozen, plausibly at a low enough value to make the Born approximation meaningful. The $\lqcd$ scale then cannot originate from renormalization. At Born level the only possibility is a non-vanishing boundary condition in solving the classical field equations for the gluon field. A conceivable scenario for hadrons is that the Born term includes the relevant non-perturbative effects and the perturbative corrections are moderate \cite{Hoyer:2016aew}.

\end{document}